\documentclass[12pt]{article}
\usepackage{graphicx}
\usepackage{epstopdf}
\usepackage{amssymb,amsmath,epsfig}

\begin{document}

\title{\bf Study of polytropes with Generalized polytropic Equation of State}

\author{M. Azam$^{1}$ \thanks{azammath@gmail.com, azam.math@ue.edu.pk},
S. A. Mardan$^2$ \thanks{syedalimardanazmi@yahoo.com, ali.azmi@umt.edu.pk},
I. Noureen$^2$ \thanks{ifra.noureen@gmail.com, ifra.noureen@umt.edu.pk} and
M. A. Rehman$^2$ \thanks{aziz3037@yahoo.com, aziz.rehman@umt.edu.pk}\\
$^1$ Division of Science and Technology, University of Education,\\
Township Campus, Lahore-54590, Pakistan.\\
$^2$ Department of Mathematics,\\ University of the Management and Technology,\\
C-II, Johar Town, Lahore-54590, Pakistan.}

\date{}

\maketitle
\begin{abstract}
The aim of this paper is to discuss the theory of Newtonian and
relativistic polytropes with generalized polytropic equation of
state. For this purpose, we formulated the general framework to
discuss the physical properties of polytrops with anisotropic inner
fluid distribution under conformally flat condition in the presence
of charge. We investigate the stability of these polytrops in the
vicinity of generalized polytropic equation through Tolman-mass. It
is concluded that one of the derived models is physically acceptable.
\end{abstract}

\section{Introduction}

The theory of polytropes have a significant role in studying the
inner structure of astrophysical compact objects (CO) very
precisely. In Newtonian gravity, various useful physical phenomena
have been addressed with polytropic equations of state (EoS).
Chandrasekhar \cite{N1} presented the basic theory of Newtonian polytropes
emerged through laws of thermodynamics for polytropic sphere.
Tooper \cite{1} provided the basic formalism of polytropes for
compressible fluid under the assumption of quasi-static
equilibrium. He extended his work for adiabatic fluid sphere and
provides the fundamental framework to derive LEe for
relativistic polytropes.
Kovetz \cite{N2} redefined some anomalies in the theory of slowly
rotating polytropes presented by  Chandrasekhar \cite{N1}.
Abramowicz \cite{N3} extended the general form of Lane-Emden equation (LEe) for spherical, planar and
cylindrical polytropes for higher dimensional spaces.

In general relativity (GR), polytropes have been
discussed by many researchers by means of LEe
which can be derived from the hydrostatic
equilibrium configuration of CO.
Cosenza et al. \cite{4} presented a heuristic procedure to construct
anisotropic models in GR.
Herrera and Santos \cite{rev1} provided a comprehensive
study to discuss the possible causes for the existence of local anisotropy
in self-gravitating systems both in Newtonian and relativistic regime.
Herrera and Barreto \cite{3}
investigated relativistic polytropes under the assumption of
post-quasi-static regime and presented a new way to describe various
physical variables like pressure, mass and energy density by means
of effective variables. They considered two possible configuration of
polytropes in the frame of GR and found that only one was physically
viable. Anisotropy plays a very vital role in the theory of GR to discuss
spherical CO.
Herrera and his coworkers \cite{rev2} developed a full set of governing equations
for spherically symmetric dissipative fluids with anisotropic stresses
to study self-gravitating systems in the framework of GR.
Herrera and Barreto \cite{5} used the
Tolman mass (measure of active gravitational mass) to find out
stability of both Newtonian and relativistic polytropes with
anisotropic inner matter configuration. Herrera et al. \cite{7}
analyzed in detail anisotropic polytropes with conformally flat
condition which was useful in reducing the parameters involved in
relativistic modified LEe. Recently, Herrera et al. \cite{77} discussed
the stability of anisotropic polytropes by means of cracking.

The presence of charge on the stars is an important physical ingredient which has a
fabulous character in studying the dynamics of astrophysical
objects. To discuss the effect of charge on CO is always a great
value in GR.
Bekenstein \cite{11} introduced the idea of hydrostatic equilibrium
to observe gravitational collapse in charged CO.
Bonnor \cite{8} observed that electric repulsion can
delay the gravitational collapse in spherically symmetric CO. Bondi has
\cite{10} provided detailed explanation of the contraction in
isotropic radiating CO with the help of Minkowski coordinates.
Koppar et al. \cite{111} provided a new technique to derive a charged generalization of known static fluid solution for spherical symmetry.
Ray et al. \cite{15} examined that high density CO can hold the amount of charge
approximately $ 10^{20}$ coulomb.
Herrera et al. \cite{newc} analyzed charged dissipative spherical fluids.
The physical meaning of structure scalars is analyzed for charged dissipative spherical fluids and for neutral dust in the presence of cosmological constant.
Takisa and Maharaj \cite{16}
studied the models of charged polytropic spherical symmetries.
Sharif and Sadiq \cite{17} presented general formalism to
study the effects of charge on anisotropic polytropes
and developed modified LEe.
Azam et al.
\cite{12} analyzed different charged CO models to
formulate their stability in the scenario of linear and quadratic
regimes. It is found that stability of these objects depend on
the electromagnetic field as well as choice of EoS.

In the development of CO models the selection of EoS is very crucial
issue. Polytropic EoS, $P_r=K \rho_0^{1+\frac{1}{n}}$ has been
used by many researchers [4-12] 
for development and
discussion of spherical CO. Chavanis \cite{18} proposed generalized polytropic EoS
$P_r=\alpha_1\rho_o+K\rho_o^{1+\frac{1}{n}}$ to discuss various
cosmological aspects of the universe. He developed the model
of early universe to elaborate the transformation from
a pre-radiation era to the radiation era for positive indices $n>0$.
In this continuity, he also produced models which described
the late universe by considering negative indices in generalized polytropic EoS \cite{19}.
Freitas and Goncalves \cite{20} used generalized polytropic EoS to
study primordial quantum fluctuations and
build a universe with constant density at the origin.
In this work, we will use generalized polytropic EoS to discuss
the theory of Newtonian polytropes and relativistic charged polytropes.

The plan of this paper is as follows. In section \textbf{2}, we shall
provide basics of Einstein-Maxwell field equation and hydrostatic
equilibrium equation. In section \textbf{3}, Newtonian polytropes
will be discussed and section \textbf{4} is devoted to observe
relativistic polytropes. Energy conditions
and conformally flat condition are discussed in section \textbf{5}.
The section \textbf{6} is devoted for stability analysis of polytropes.
In the last section, we shall conclude our results.

\section{The Einstein-Maxwell Field Equations}

We consider static spherically symmetric space time
\begin{equation}\label{1}
ds^2=e^{\nu}dt^{2}-e^{\lambda}dr^{2}-r^2 d\theta^{2}-r^2 \sin^2\theta{d\phi^2},
\end{equation}
where $\nu,~\lambda$ are functions of $r$ only. The energy-momentum
tensor for anisotropic matter distribution is given by
\begin{equation}\label{2}
T_{i j}=(P_t+\rho) V_{i} V_{j} -g_{i j}P_t +(P_r - P_t) S_{i}
S_{j},
\end{equation}
where  $P_t,~P_r$ and $\rho$ respectively represent the tangential pressure,
radial pressure and energy density of the inner fluid distribution. The four velocity
$V_{i}$ and four vector $S_{i}$ satisfying the following equations
\begin{equation}\label{3}
V_{i} = e^{\frac{\nu}{2}} \delta^0_i,~~~V^{i}V_{i} = 1,~~~ S_{i} =
e^{\frac{\lambda}{2}} \delta^1_i,~~~S^{i}S_{i} =-1, ~~~S^{i}V_{i} =0.
\end{equation}
The electromagnetic energy-momentum tensor is defined by
\begin{equation}\label{4}
T^{(em)}_{ij}=\frac{1}{4\pi}(F_{i}^{m} F_{j m}-\frac{1}{4} F^{mn} F_{mn} g_{ij}),
\end{equation}
where
\begin{equation}\label{5}
F_{ij}=\psi_{j,i}-\psi_{i,j} ,
\end{equation}
is the Maxwell field tensor and satisfies the following relation
\begin{equation}\label{6}
F^{ij}_{;j}=\mu_0 J^i,~~~F_{[ij;k]}=0,
\end{equation}
where $\psi_i$ is the four-potential, $\mu_0$ is the magnetic
permeability and $J^i$ is the four-current.
Also, four-potential and four-velocity satisfies the following relations
in co-moving coordinates
\begin{equation}\label{7}
\psi_{i}=\psi(r) \delta^0_i,~~~J^i=\sigma V^i,~~~~~~ i=0,1,2,3,
\end{equation}
where $\psi$ is the scalar potential and $\sigma$ is the charge
density. The Maxwell equation (6) yields
\begin{equation}\label{8}
\psi^{\prime\prime}+\Big(\frac{2}{r}-\frac{\nu^\prime}{2}
-\frac{\lambda^\prime}{2}\Big)\psi^\prime=4 \pi
 \sigma e^{\frac{\nu+\lambda}{2}},
\end{equation}
where $``\prime"$ denotes the differentiation with respect to $r$. From above, we have
\begin{equation}\label{9}
\psi^{\prime}=\frac{q(r)}{r^2}e^{\frac{\nu+\lambda}{2}}.
\end{equation}
where $q(r)=4 \pi \int_0^r \mu e^{\frac{\lambda}{2}} r^2 dr $
represents the total charge inside the sphere.

The Einstein-Maxwell field equations for line element Eq.$(\ref{1})$
are given by
\begin{eqnarray}\label{10}
\frac{\lambda^\prime
e^{-\lambda}}{r}+\frac{(1-e^{-\lambda})}{r^2}=8\pi \rho
-\frac{q^2}{r^4}, \\\label{11}
\frac{\nu^\prime e^{-\lambda}}{r}-\frac{(1-e^{-\lambda})}{r^2}=8\pi P_r
-\frac{q^2}{r^4}, \\\label{12}
e^{-\lambda} \bigg[\frac{\nu^{\prime\prime}}{2}-\frac{\nu^\prime
\lambda^\prime}{4}+\frac{\nu^{\prime^2}}{4}+\frac{\lambda^\prime
-\nu^\prime}{2r}\bigg] = 8 \pi P_t+\frac{q^2}{r^4},
\end{eqnarray}
solving Eqs.$(\ref{10})$-$(\ref{12})$ simultaneously lead to
hydrostatic equilibrium equation \cite{17}
\begin{equation}\label{13}
\frac{d P_r}{dr}-\frac{2}{r}\Big(\Delta+\frac{q q\prime}
{8\pi r^3}\Big)+\frac{(4\pi r^4 P_r-q^2+m r)}{r(r^2-2mr+q^2)}=0,
\end{equation}
where we have used $\Delta=(P_t-P_r)$.

The junction conditions are commonly used to relate
inner and outer regions of CO over a boundary surface. The
choice of outer geometry totally relies on the inner fluid distribution of CO.
For spherical symmetry, if matter in the interior of a star is charged anisotropic,
we take the Reissner-Nordstr\"{a}m space-time as the exterior geometry
\begin{equation}\label{24}
ds^2=\Big(1-\frac{2M}{r}+\frac{Q^2}{r^2}\Big)dt^2-
\Big(1-\frac{2M}{r}+\frac{Q^2}{r^2}\Big)^{-1}dr^2-r^2
d\theta^{2}-r^2 \sin^2\theta{d\phi^2}.
\end{equation}
Junction conditions have very significant role in the study of
CO. These conditions elaborates whether the combination of two space-time metrics
provide a physically viable solution or not when a hyper-surface
divides the space-time into interior and exterior regions.
The smooth matching of interior and exterior regions through first
and second fundamental forms [24-26]
yields the following relations on
the boundary surface, i.e.,
\begin{equation}\label{25}
e^{\nu}
=e^{-\lambda}=\Big(1-\frac{2M}{r}+\frac{Q^2}{r^2}\Big),~~m(r)=M,~
q(r)=Q,~P_r=0,
\end{equation}
and Misner-Sharp mass leads to \cite{M1,M2}
\begin{equation}\label{14}
m(r)=\frac{r}{2}(1-e^{-\lambda}+\frac{q^2}{r^2}).
\end{equation}
In spherical symmetry the Misner-Sharp mass can be described
as a measure of the total energy within a sphere of a radius
$r$ at a time $t$.
The theory of polytropes is based on the assumption of hydrostatic equilibrium and
polytropic EoS. In this work, we discuss Newtonian and relativistic
ploytropes by using generalized polytropic EoS, which is the
combination of linear and polytropic EoS.

\section{Newtonian Polytrope}

In the framework of Newtonian gravity, the polytropic
EoS are very useful in describing a huge
variety of situations like inner pressure, types of inner fluid distribution etc.
In this section, we formulate Newtonian polytropes through
hydrostatic equilibrium equation in the scenario of generalized
polytropic equation of state. Polytropes are supposed to be in
hydrostatic equilibrium, we have to keep in mind that the appearance of cracking
occurs when the system is taken out of equilibrium. We consider the following basic equations \cite{3}
\begin{equation}\label{15}
\frac{d P_r}{dr}=-\frac{d\phi}{dr}\rho_o,
\end{equation}
and in spherical coordinates the Poisson equation is given by
\begin{equation}\label{16}
\frac{1}{r^2}\frac{d}{dr}\Big(r^2 \frac{d \phi}{dr}\Big)=4\pi\rho_o,
\end{equation}
here $\rho_o$ is the mass density and $\phi$ is taken to be
Newtonian gravitational potential. The above
Eqs.$(\ref{15})$-$(\ref{16})$ along with the
generalized ploytropic EoS \cite{18,19}
\begin{equation}\label{17}
P_r=\alpha_1\rho_o+K\rho_o^{\gamma}=\alpha_1\rho_o+
K\rho_o^{1+\frac{1}{n}},
\end{equation}
leads to modified LEe (for $ \gamma \neq 1$). Here
$K$ is the polytropic constant and $n$ is the
polytropic index. We take
\begin{equation}\label{18}
\rho_o=\rho_{gc}\theta^n(r),
\end{equation}
where $\rho_{gc}$ is the density evaluated at center $c$,
using Eq.$(\ref{17})$-$(\ref{18})$ in Eq.$(\ref{15})$, we get
\begin{equation}\label{19}
\frac{n\alpha_1
\theta^\prime}{\theta}+(n+1)K\rho_{gc}\theta^\prime=-\frac{d\phi}{dr}.
\end{equation}
From Eqs. $(\ref{16})$, $(\ref{18})$ and $(\ref{19})$, we obtain
\begin{equation}\label{20}
-A^2_3\Big(\frac{r}{A_2}\Big)^{-2}\frac{d}{dr}
\Big(\frac{r^2 \theta^\prime}{\theta}\Big)
-\Big(\frac{r}{A_2}\Big)^{-2}\frac{d}{dr}
\Big(r^2 \theta^\prime\Big)=\theta^n,
\end{equation}
where $A^2_2$ and $ A^2_3$ are given by
\begin{equation}\label{21}
A^2_2 = \frac{4\pi \rho_{gc}^{1-\frac{1}{n}}}{K(n+1)},~~~A^2_3 =
\frac{Kn(n+1)\alpha_1}{16 \pi^2 \rho_{gc}^{2-\frac{1}{n}}}.
\end{equation}
Inserting $\xi=\frac{r}{A_2}$ in above equation, we obtained the
modified form of LEe
\begin{equation}\label{22}
\Big(\frac{A^2_3+\theta}{\theta}\Big)\frac{d^2 \theta}
{d \xi^2}+\frac{2}{\xi}\Big(\frac{A^2_3+\theta}
{\theta}\Big)\frac{d \theta}{d \xi}-\Big(\frac{A^2_3}
{\theta^2}\Big)\Big(\frac{d \theta}{d \xi}\Big)^2+\theta^n=0,
\end{equation}
with the boundary conditions
\begin{equation}\label{23}
\frac{d \theta}{d \xi}(\xi=0)=0,~~~\theta(\xi=0)=1.
\end{equation}
The boundary of sphere is defined by
$\xi=\xi_n$, such that $\theta(\xi_n)=0$.

\section{The Relativistic Polytropes}

This section deals with the relativistic configuration of polytropes with
generalized EoS. The generalized polytropic EoS is the linear combination
of linear EoS  $``P_r=\alpha_1\rho_o" $ and polytropic EoS
$``P_r=K\rho_o^{1+\frac{1}{n}}"$. The linear EoS describes
pressureless $(\alpha_1=0)$ or radiation $(\alpha_1=\frac{1}{3})$ matter.
The polytropic part describes cosmology of the early universe for positive values
of polytropic index whereas it elaborates the late time universe with negative values
of polytropic index \cite{18,19}.
In order to discuss the cosmic behavior, $``\rho_o"$ was taken to be Planck density
but for relativistic discussion we will consider it as mass density and
total energy density in case \textbf{1} and case \textbf{2} respectively.
Here, we shall present the general formalism for relativistic polytropes with generalized polytropic EoS
and investigate the presence of charge in the following two cases.

\subsection{Case 1}

Here, we consider the generalized ploytropic EoS as
\begin{equation}\label{26}
P_r=\alpha_1\rho_o+K\rho_o^{\gamma}=\alpha_1\rho_o+K\rho_o^{1+\frac{1}{n}},
\end{equation}
so that the original polytropic part remain conserved,
also the mass density $\rho_o$ is related to total energy density
$\rho$ as \cite{7}
\begin{equation}\label{27}
\rho=\rho_{o}+n P_r.
\end{equation}
Now taking following assumptions
\begin{equation*}
\alpha=\frac{P_{rc}}{\rho_{gc}},~~~\alpha_2
=1+(n+1)(\alpha_1+\alpha\theta),~~~\alpha_3
=(n+1)\alpha,~~~\alpha_4=\frac{4\pi P_{rc} q^2}{\alpha \alpha_3},
\end{equation*}
\begin{equation}\label{28}
r=\frac{\xi}{A},~~~\rho_{o}=\rho_{gc}\theta^n,~~~m(r)
=\frac{4\pi\rho_{gc} v({\xi})}{A^3},~~~A^2=
\frac{4\pi\rho_{gc}}{(n+1)\alpha},
\end{equation}
where $P_{rc}$ is the pressure at center of the star,
$\rho_{gc}$ is the mass density
evaluated at the center of CO, $\xi$, $\theta$ and
$v$ are dimensionless variables.
Using above assumptions along with EoS
$(\ref{26})$, the hydrostatic equilibrium equation Eq.$(\ref{13})$ implies
\begin{eqnarray}\label{29} \notag
&&\Big(1-2\alpha_3 \frac{v(\xi)}{\xi}+\frac{\alpha_4}
{\xi^2} \Big)\Big(\frac{n\alpha_1 \theta^{-1}
+\alpha_3}{\alpha_2\alpha_3} \xi^2 \frac{d\theta}{d\xi}
-2\frac{\alpha^2\alpha^2_3\xi^3\Delta+2\pi P^2_{rc} q \frac{d q}{d\xi}}
{\alpha \alpha_2 \alpha_3^3 P_{rc}\xi^2}\theta^{-n}\Big)\\
&&- \frac{\alpha_4}{\alpha_3
\xi}+v(\xi)+(\alpha_1+\alpha\theta)\xi^3\theta^n=0.
\end{eqnarray}
Now differentiating Eq.$(\ref{14})$ with respect to $``r"$ and using the
assumptions given in Eq.$(\ref{28})$, we get
\begin{eqnarray}\label{30}
\frac{dv(\xi)}{d\xi}=\xi^2\theta^n(1+n\alpha_1+n\alpha\theta)
-\frac{\alpha_4}{\alpha_3\xi^2}+\frac{\alpha_4}
{\alpha_3 \xi q}\frac{d q}{d\xi}.
\end{eqnarray}
Thus Eq.$(\ref{29})$ coupled with Eq.$(\ref{30})$
yields modified LEe (see appendix Eq.$(\ref{101})$)
which describe the relativistic polytropes in the presence of charge
$q$ with generalized polytropic EoS.

\subsection{Case 2}

Here, we consider the generalized ploytropic EoS as
\begin{equation}\label{31}
P_r=\alpha_1\rho+K\rho^{1+\frac{1}{n}},
\end{equation}
where mass density $\rho_o$ is replaced by
total energy density $\rho$ in Eq.$(\ref{26})$ and
they are related to each other as \cite{7}
\begin{equation}
\rho=\frac{\rho_o}{\big(1-K \rho_o^{\frac{1}{n}}\big)^n}.
\end{equation}
We take following assumptions
\begin{equation*}
\alpha=\frac{P_{rc}}{\rho_{c}},~~~\alpha_5=1+\alpha_1+\alpha\theta,
\end{equation*}
\begin{equation}\label{32}
r=\frac{\xi}{A},~~~\rho_{o}=\rho_{c}\theta^n,~~~m(r)
=\frac{4\pi\rho_{c}
v({\xi})}{A^3},~~~A^2=\frac{4\pi\rho_{c}}{(n+1)\alpha},
\end{equation}
where $c$ represents the quantity at center of the star,
$\alpha_2$, $\alpha_3$ and $\alpha_4$
 are same expressions as in Eq.$(\ref{28})$
with $\alpha$ defined in Eq.$(\ref{32})$.
Using above assumptions along with EoS $(\ref{31})$,
the hydrostatic equilibrium equation $(\ref{13})$ becomes
\begin{eqnarray}\label{33} \notag
&&\Big(1-2\alpha_3 \frac{v(\xi)}{\xi}+\frac{\alpha_4}
{\xi^2} \Big)\Big(\frac{n\alpha_1 \theta^{-1}
+\alpha_3}{\alpha_3\alpha_5} \xi^2
\frac{d\theta}{d\xi}-2\frac{\alpha^2\alpha^2_3\xi^3\Delta
+2\pi P^2_{rc} q \frac{d q}{d\xi}}
{\alpha \alpha_3^3 \alpha_5 P_{rc}\xi^2}\theta^{-n}\Big)\\
&&- \frac{\alpha_4}{\alpha_3
\xi}+v(\xi)+(\alpha_1+\alpha\theta)\xi^3\theta^n=0.
\end{eqnarray}
Now differentiating Eq.$(\ref{14})$ with respect to
$``r"$ and using assumptions given in Eq.$(\ref{32})$, we get
\begin{eqnarray}\label{34}
\frac{dv(\xi)}{d\xi}=\xi^2\theta^n-\frac{\alpha_4}{\alpha_3
\xi^2}+\frac{\alpha_4}{\alpha_3 \xi q}\frac{d q}{d\xi}.
\end{eqnarray}
Eq.$(\ref{33})$ coupled with Eq.$(\ref{34})$
gives modified LEe (see appendix Eq.$(\ref{102})$) represent the relativistic
polytropes in the presence of charge $q$ with generalized polytropic EoS.

\section{Energy Conditions and Conformally Flat Condition}

The energy conditions in GR are designed to
obtain maximum possible information without enforcing a particular EoS.
The energy conditions are provided with the sense that energy
density cannot be negative because if it allows
the random $+ve$ and $-ve$ energy regions,
the empty space would become unstable.
The energy conditions satisfied by all the spherically symmetric models are \cite{17,En1}
\begin{equation}\label{35}
\rho+\frac{q^2}{8\pi r^4}>0,~~~\frac{P_{r}}{\rho}
\leq 1+\frac{q^2}{4\pi \rho r^4},~~~\frac{P_{t}}{\rho}\leq 1.
\end{equation}
For case \textbf{1} the conditions given in Eq.$(\ref{35})$ turn out to be
\begin{equation*}
1+n(\alpha+\alpha_1)\theta+\frac{\alpha_4 \theta^{-n}}
{2\alpha_3 \xi^4}>0,~~~1\leq n
+(\alpha_1+\alpha\theta)^{-1}+\frac{\alpha_4 \theta^{-n}}
{(\alpha_1+\alpha\theta)\alpha_3 \xi^4},
\end{equation*}
\begin{equation}\label{36}
\frac{3 v(\xi)}{\xi^3}+\frac{2\alpha_4\frac{d q}{d \xi}}{\alpha_3 q \xi^3}
+\alpha_1 \theta+\alpha\theta^{n+1}\leq \frac{4\alpha_4}{\alpha_3 \xi^3},
\end{equation}
and for case \textbf{2} the conditions given in Eq.$(\ref{35})$ emerge as
\begin{equation}\label{37}
1+\frac{\alpha_4 \theta^{-n}}{\alpha_3 \xi^4}>0,~\alpha_1+\alpha\theta
-\frac{\alpha_4 \theta^{-n}}{\alpha_3 \xi^4}\leq 1,~\frac{3 v(\xi)}{\xi^3}
+\frac{2\alpha_4\frac{d q}{d \xi}}{\alpha_3 q \xi^3}+\alpha_1
+\alpha\theta^{n+1}\leq \frac{4\alpha_4}{\alpha_3 \xi^4}.
\end{equation}

We observe that coupled Eqs.$(\ref{29})$-$(\ref{30})$
and Eqs.$(\ref{33})$-$(\ref{34})$ form a system of differential equations.
These systems involve three variables
and we want some additional information to study polytropic CO.
We use conformally flat condition to
reduce one variable in the above said system of equations.
The electric part of Weyl tensor is related to Weyl scalar given by \cite{7,17}
\begin{equation}\label{38}
W=\frac{r^3 e^{-\lambda}}{6}\Bigg(\frac{e^\lambda}{r^2}
+\frac{\lambda^\prime \nu^\prime}{4}
-\frac{1}{r^2} -\frac{\nu^{\prime 2}}{4}-
\frac{\nu^{\prime\prime}}{2}-\frac{\lambda^\prime \nu^\prime}{2 r}\Bigg).
\end{equation}
Now using conformally flat condition, i.e., $W=0$, along with field
equations Eqs.$(\ref{10})$-$(\ref{12})$ in equation Eq.$(\ref{38})$, we get
\begin{equation}\label{39}
\Delta=P_t-P_r=\frac{e^{-\lambda}}{4\pi}\Bigg(\frac{e^{\lambda}}{r^2}
-\frac{\lambda^\prime}{2r} -\frac{1}{r^2}\Bigg)-\frac{q^2}{4 \pi r^4}.
\end{equation}
The above equation along with Eq.$(\ref{28})$ and Eq.$(\ref{30})$ for case \textbf{1} yields
\begin{equation}\label{40}
\Delta=\rho_{gc}\Bigg((1+n\alpha_1+n\alpha\theta)
\theta^n+3\frac{v(\xi)}{\xi^3}
-4\frac{\alpha_4}{\alpha_3 \xi^4}+2
\frac{\alpha_4 \frac{d q}{d \xi}}{\alpha_3 q \xi^3}\Bigg).
\end{equation}
Similarly, anisotropy parameter for case \textbf{2} turn out to be
\begin{equation}\label{41}
\Delta=\rho_{c}\Bigg(\theta^n+3\frac{v(\xi)}{\xi^3}
-4\frac{\alpha_4}{\alpha_3 \xi^4}+2\frac{\alpha_4 \frac{d q}{d
\xi}}{\alpha_3 q \xi^3}\Bigg).
\end{equation}
Using Eq.$(\ref{40})$ in Eq.$(\ref{29})$, we obtain first equation of
coupled differential system corresponding to case \textbf{1}
\begin{eqnarray}\label{42} \notag
&&\Big(1-2\alpha_3 \frac{v(\xi)}{\xi^2}
+\frac{\alpha_4}{\xi^2} \Big)
\Big(\frac{n\alpha_1 \theta^{-1}+\alpha_3}
{\alpha_2\alpha_3} \xi^2 \frac{d\theta}{d\xi}
-\frac{2\xi}{\alpha_2\alpha_3}
(1+n\alpha_1+n \alpha \theta)+\\\notag
&&\Big(-\frac{6\alpha v(\xi)}{\alpha_2\alpha_3 P_{rc}\xi^2}
+\frac{8\alpha\alpha_4}{\alpha_2\alpha_3^2 P_{rc} \xi^3}
-\frac{4\alpha\alpha_4 \frac{d q}{d \xi}}{\alpha_2\alpha_3^2}
+\frac{4\pi P^2_{rc} q \frac{d q}{d \xi}}
{\alpha\alpha_2\alpha_3^3 \xi^2}\Big)\theta^{-n}\Big)\\
&&-\frac{\alpha_4}{\alpha_3
\xi}+v(\xi)+(\alpha_1+\alpha\theta)\xi^3\theta^n=0.
\end{eqnarray}
The above equation coupled with Eq.$(\ref{30})$
provides modified LEe  (see appendix Eq.$(\ref{103})$)
which describe the conformally flat polytrope for case \textbf{1}.\\
In the same way, for case \textbf{2}, using Eq.$(\ref{40})$ in Eq.$(\ref{33})$, we obtained
\begin{eqnarray}\label{43} \notag
&&\Big(1-2\alpha_3 \frac{v(\xi)}{\xi}+\frac{\alpha_4}{\xi^2} \Big)
\Big(\frac{n\alpha_1 \theta^{-1}+\alpha_3}
{\alpha_3\alpha_5} \xi^2 \frac{d\theta}{d\xi}
-\frac{2\xi}{\alpha_3\alpha_5}
+\Big(-\frac{6 v(\xi)}{\alpha_3\alpha_5 P_{rc}\xi^2}\\\notag
&&+\frac{8\alpha_4}{\alpha_3\alpha_5 P_{rc} \xi^3}
-\frac{4\alpha_4 \frac{d q}{d \xi}}{\alpha_3^2\alpha_5q \xi^2}
-\frac{4\pi P_{rc} q \frac{d q}{d \xi}}
{\alpha\alpha_3^3 \xi^2}\Big)\theta^{-n}\Big)\\
&&-\frac{\alpha_4}{\alpha_3
\xi}+v(\xi)+(\alpha_1+\alpha\theta)\xi^3\theta^n=0.
\end{eqnarray}
The above equation coupled with Eq.$(\ref{34})$
leads to modified LEe  (see appendix Eq.$(\ref{104})$)
for conformally flat polytropes (case \textbf{2}).

\section{Stability Analysis}

The stability of the model can be discussed by
means of Tolman mass, which measure the
active gravitational mass of CO. The important feature of Tolman mass is that it can be
evaluated by integrating over the region occupied
by matter or electromagnetic energy [30-32]. 
The modified form of Tolman mass for
anisotropic spherically symmetric metric $(\ref{1})$ is given by \cite{17}
\begin{eqnarray}\label{44}
m_T=M\frac{r_\Sigma ^3}{r^3}+r^3 \int_r^{r_\Sigma}
e^{(\lambda+\nu)/2}
\Big(\frac{2}{\tilde{r}^4}W-\frac{4\pi\Delta}
{\tilde{r}}+\frac{q^2}{16\tilde{r}^5}
\Big)d\tilde{r},
\end{eqnarray}
where $\Sigma$ represents the values calculated at the boundary of CO.
In order to get an expression for $\nu$, we will solve
Einstein-Maxwell field Eqs.$(\ref{10})$-$(\ref{12})$ simultaneously, we have
\begin{eqnarray}\label{45}
\frac{\nu^\prime}{2}=\frac{(4\pi r^4 P_r-q^2+m r)}
{r(r^2-2mr+q^2)}.
\end{eqnarray}
The integration of above equation yields
\begin{eqnarray}\label{46}
\nu=\nu_{\Sigma}-\int_r^{r_\Sigma}2
\frac{(4\pi r^4 P_r-q^2+m r)}{r(r^2-2mr+q^2)}dr.
\end{eqnarray}
We define the dimensionless variables as
\begin{equation}\label{47}
x=\frac{r}{r_\Sigma}=\frac{\xi}{\tilde{A}},~~~\tilde{A}=A
r_\Sigma,~~~y=\frac{M}{r_\Sigma},~~~\tilde{m}=\frac{m}{M}.
\end{equation}
For case \textbf{1}, using Eqs.$(\ref{28})$,
$(\ref{45})$ and $(\ref{47})$ in Eq.$(\ref{44})$, we get
\begin{eqnarray}\label{48}\notag
\frac{m_T}{M}&=& x^3+ \frac{(n+1)\alpha x^3 \tilde{A}^2}{4 \pi
y}\int_x^1 \Big(\frac{1-2y+Q^2/r_\Sigma^2}{1-2(n+1)\alpha
v/x\tilde{A}+q^2/x^2 r_\Sigma^2}\Big)^{1/2}\\ \notag &&\times
exp\Big(\int_x^1 -\frac{4\pi (\alpha_1+\alpha\theta) \theta^{n} x^4
r_\Sigma^4-q^2(n+1)\alpha v x r_\Sigma^2/\tilde{A}} {r_\Sigma^3 x^4
-2(n+1) \alpha v r_\Sigma^4 x^3 /\tilde{A}+ q^2 r_\Sigma^2
x}\Big)dx\\  &&\times\Big(\frac{-4\pi \Omega}{x
r_\Sigma}+\frac{q^2}{16 x^5 r_\Sigma \rho_{gc}}\Big)dx,
\end{eqnarray}
where $y=\frac{(n+1)\alpha \nu_\Sigma}{\xi_\Sigma}$ and $\Omega$ is given by
\begin{equation}\label{49}
\Omega=\frac{\Delta}{\rho_{gc}}=\Bigg((1+n\alpha_1+n\alpha\theta)
\theta^n+3\frac{v(\xi)}{\xi^3}-4\frac{\alpha_4}{\alpha_3
\xi^4}+2\frac{\alpha_4 \frac{d q}{d \xi}}{\alpha_3 q \xi^3}\Bigg).
\end{equation}
For case \textbf{2}, inserting Eqs.$(\ref{32})$, $(\ref{45})$ and
$(\ref{47})$ in Eq.$(\ref{44})$, we obtain
\begin{eqnarray}\label{50}\notag
\frac{m_T}{M}&=& x^3+ \frac{(n+1)\alpha x^3 \tilde{A}^2}{4 \pi y}\int_x^1 \Big(\frac{1-2y+Q^2/r_\Sigma^2}{1-2(n+1)\alpha v/x\tilde{A}+q^2/x^2 r_\Sigma^2}\Big)^{1/2}\\ \notag &&\times
exp\Big(\int_x^1 -\frac{4\pi (\alpha_1+\alpha\theta)\theta^{n} x^4 r_\Sigma^4-q^2(n+1)\alpha v x r_\Sigma^2/\tilde{A}}{r_\Sigma^3 x^4
-2(n+1) \alpha v r_\Sigma^4 x^3 /\tilde{A}+ q^2 r_\Sigma^2 x}\Big)dx\\
&&\times
\Big(\frac{-4\pi \Omega}{x
r_\Sigma}+\frac{q^2}{16 x^5 r_\Sigma \rho_{c}}\Big)dx,
\end{eqnarray}
and $\Omega$ is given by
\begin{equation}\label{51}
\Omega=\frac{\Delta}{\rho_{gc}}=\Bigg
(\theta^n+3\frac{v(\xi)}{\xi^3}-4\frac{\alpha_4}{\alpha_3
\xi^4}+2\frac{\alpha_4 \frac{d q}{d \xi}}{\alpha_3 q \xi^3}\Bigg).
\end{equation}
\begin{figure}
\centering
\includegraphics[width=80mm]{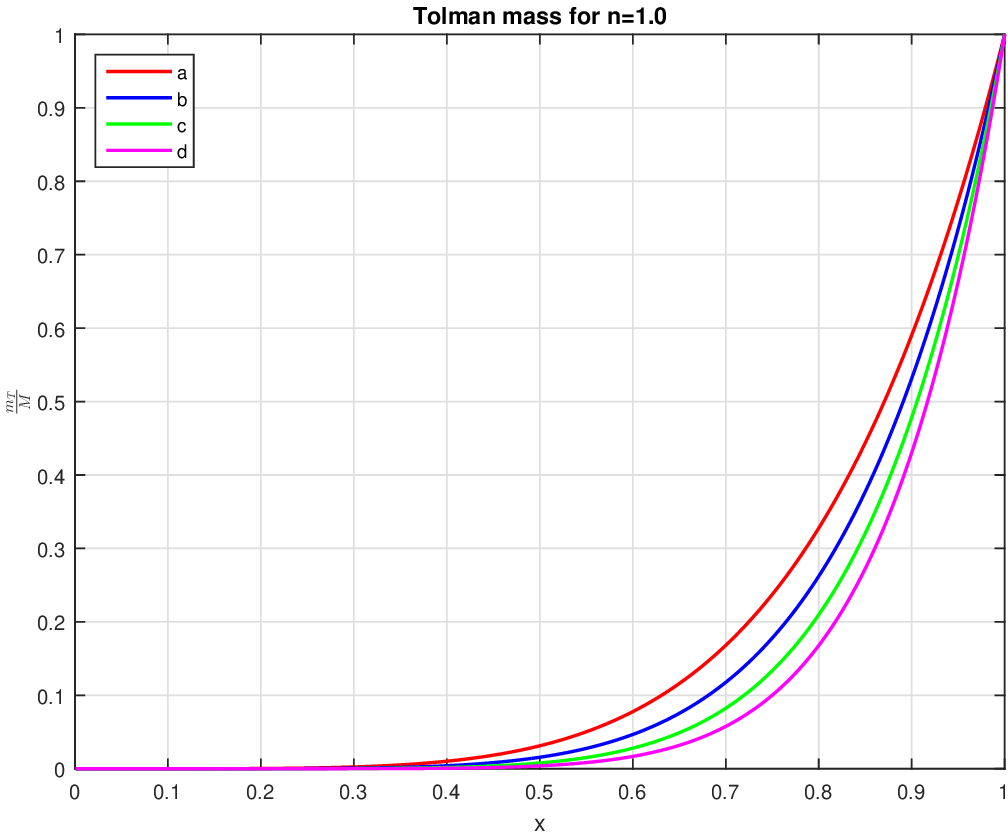}
\caption{Case $1$: $\frac{m_T}{M}$ as a function of $x$ for $n=1$,
curve $a$: $\alpha=8\times 10^{-11}$,~~y=0.3991, Q=0.2 $M_\odot$,
curve $b$: $\alpha=10^{-10}$,~~y=0.4091, Q=0.4 $M_\odot$, curve $c$:
$\alpha=2 \times 10^{-10}$,~~y=0.3998, Q=0.6 $M_\odot$, curve $d$:
$\alpha=4 \times 10^{-10}$,~~y=0.3858, Q=0.64 $M_\odot$.}.
\end{figure}
\begin{figure}
\centering
\includegraphics[width=80mm]{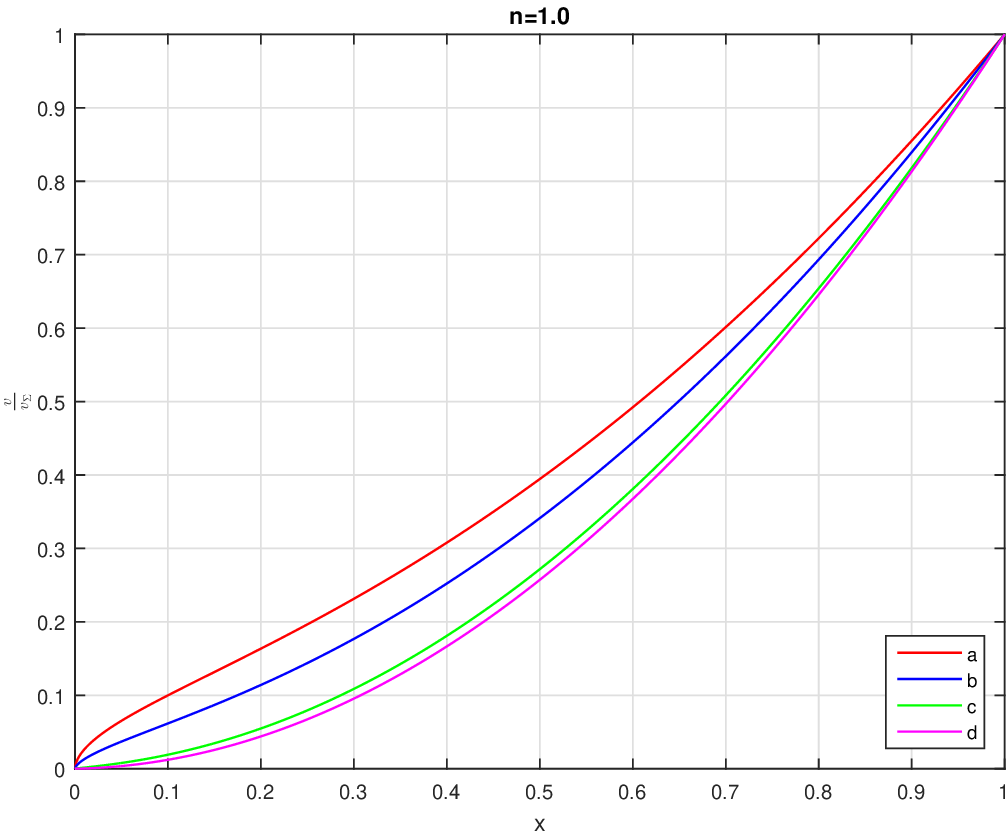}
\caption{Case $1$: $\frac{v}{\nu_\Sigma}$ as a function of $x$ for
$n=1$, curve $a$: $\alpha=8\times 10^{-11}$,~~y=0.3991, Q=0.2
$M_\odot$, curve $b$: $\alpha=10^{-10}$,~~y=0.4091, Q=0.4 $M_\odot$,
curve $c$: $\alpha=2 \times 10^{-10}$,~~y=0.3998, Q=0.6 $M_\odot$,
curve $d$: $\alpha=4 \times 10^{-10}$,~~y=0.3858, Q=0.64
$M_\odot$.}.
\end{figure}
\begin{figure}
\centering
\includegraphics[width=80mm]{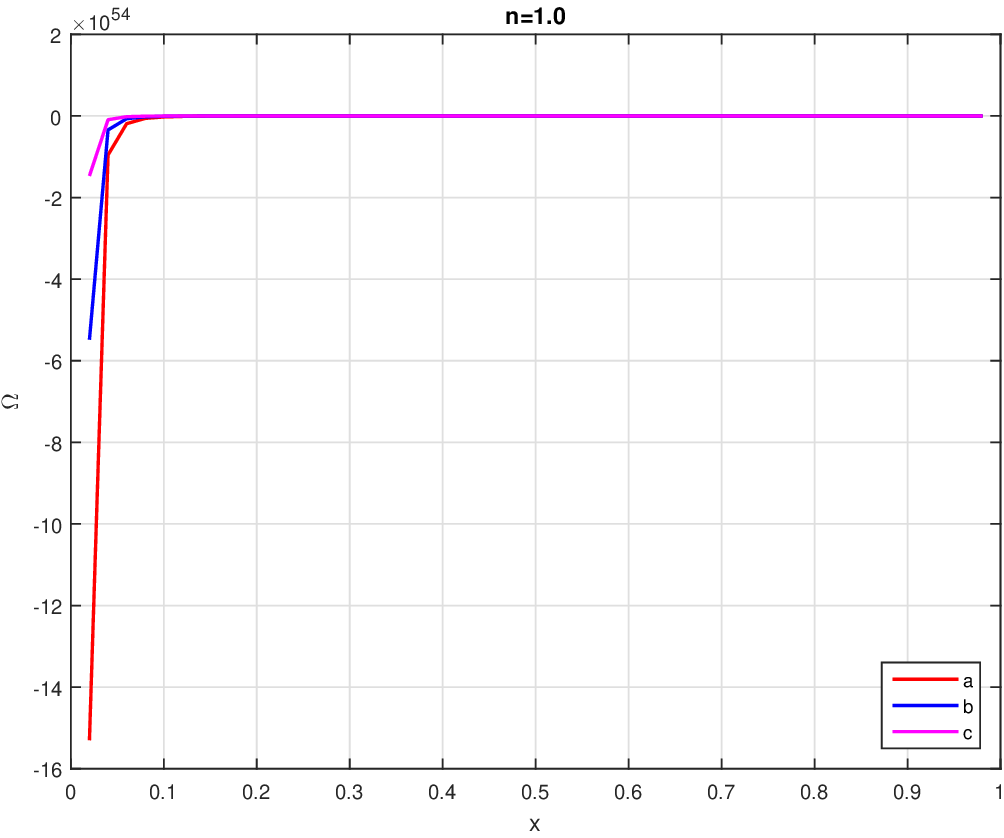}
\caption{Case $1$: $\Omega$ as a function of $x$ for $n=1$, curve
$a$: $\alpha=8\times 10^{-11}$,~~y=0.3991, Q=0.2 $M_\odot$, curve
$b$: $\alpha=10^{-10}$,~~y=0.4091, Q=0.4 $M_\odot$, curve $c$:
$\alpha=2 \times 10^{-10}$,~~y=0.3998, Q=0.6 $M_\odot$.}.
\end{figure}

In order to study the physical viability of these models, we have calculated
the Tolman mass whose behavior describes some physical features of these models, particularly
stability of the model. Figures \textbf{1-3} have been
plotted for \textbf{case 1} (relativistic polytropes) corresponding to
different parametric values $n$, $\alpha$, $y$, $Q$ and $\alpha_1=0.5$  \cite{7,17}.
Figure \textbf{1} shows that Tolman mass is gradually decreasing and does not show any
abnormal behavior as the value of $\alpha$ and charge $Q$ increases.
The curves become steeper with higher values of $\alpha$ and charge
$Q$ but they remain positive which is consistent with the
results of Herrera et al. \cite{7} for neutral polytropes.
It is noted that for smaller values of $y$, the sphere gets more
compact which corresponds to equilibrium configuration. This behavior
is shown by the migration of Tolman mass towards the boundary surface.
In terms of stability, it is happened due to sharper reduction of the active gravitational mass
in the inner regions of the sphere which may correspond to more stable configurations for
small values of $y$. The smooth behavior of Figure \textbf{2} describes the solution of Eq.$(\ref{30})$ which
shows that the stability of the model can be enhanced through decreasing the values of $y$. Figure
\textbf{3} represents the anisotropy of the model which is larger
near the center and become weaker near the surface. For \textbf{case 2}, the model does not satisfy
the energy conditions presented in Eq.$(\ref{37})$ for various values of parameters.
These results are consistent with \cite{7,17} for charge and neutral polytropes.

\section{Conclusion and Discussion}

In this work, we have developed the general framework to discuss the Newtonian
and relativistic polytropes with generalized polytropic EoS in the
presence of electromagnetic field under conformally flat condition.
The generalized polytropic EoS $P_r=\alpha_1\rho_o+K\rho_o^{1+\frac{1}{n}}$
is the union of linear and polytropic EoS.
It is widely used in cosmology to explain different era of universe
with the help of mathematical models.
In cosmic scenario $``\rho_o"$ was taken to be Planck density
but for relativistic discussion we have taken it as mass density and
total energy density in case \textbf{1} and case \textbf{2}, respectively.
In order to discuss the physical features of Newtonian
and relativistic ploytropes, we have formulated the
general formalism to obtain modified LEe.
The solutions of the LEe are called polytropes, which depend
on density profile function of dimensionless radius $\xi$ and
the order of solution is constrained with value of polytropic index $n$.
The LEe are helpful to explain relativistic CO
as they produced simple solution to describe the internal structure
of CO. But the cost of simplicity is a power law relationship
between pressure and density which should be valid
throughout the inner matter of CO. The LEe have been initially developed
for relativistic charged ploytropes with anisotropic factor involved in them
(see Eqs.$(\ref{101})$ and $(\ref{102})$), which have three unknown variables.
The conformally flat conditions is used to simplify these equations
by eliminating anisotropy factor, which corresponds to
modified LEe (see Eqs.$(\ref{103})$ and $(\ref{104})$),
for charged polytropic spheres in the context of generalized polytropic EoS.
The energy conditions are helpful in order to check the
physical viability of CO in GR without enforcing EoS. The energy conditions for both cases
of relativistic polytropes has been developed in the presence of charge.

The stability of the relativistic polytropes
is analyzed with the help of Tolman mass.
For case \textbf{1}, the model behaved well as shown in Figures
\textbf{1-3}. The model developed here are anisotropic (in pressure)
and helped us to discuss highly charged CO solutions.
Figure \textbf{1} represents the physical behavior of Tolman mass for
different values of charge and model's parameters \cite{7,17}.
The Tolman mass which
is the measure of active gravitational mass gradually shifts towards
the boundary surface but remain stable and did not presents any
fluctuations. Figure \textbf{1} represents that slow migration
of the Tolman mass towards the boundary as charge increases
corresponds to more stability of the model.
It is worth mentioning here that our results
for Tolman mass are considerably consistent with work
of Herrera et al. \cite{7} for polytropic EoS.

Figure \textbf{2} describes the solution of Eq.$(\ref{30})$ which
shows that the stability of the model can be
enhanced through decreasing the values of $y$.
This figure has been plotted for indicated
values of triplet $n,~\alpha,~Q$ and
shows stable behavior for small values of $\alpha$.
Also, the configuration of solution remain stable for increasing value of charge, although
solution curves start bending towards
the boundary of the CO as shown in Figure \textbf{2}.
It is also evident that the behavior of parameter
is disperse in the beginning i.e., near the center of CO
and it becomes more consistent as we move towards the boundary
with the increase of charge.
So, we can say charge is the stabilizing factor here
for relativistic polytropes with generalized polytropic EoS.
Thus, the discussion of charge is essential for the study
and exploration of relativistic polytropic CO for their physical viability.

The inclusion of anisotropy plays a vital role to study the
astrophysical CO.
The inner fluid distribution is assumed to be anisotropic for
spherically symmetric CO.
Figure \textbf{3} represents the graphs of $\Omega$,
which is the ratio of anisotropy and central mass density.
The curves remains stable even after increase of charge
but rapidly move towards boundary of sphere and this phenomenon shows
that anisotropy is affected by the presence of electromagnetic field.
Hence, the presence of anisotropy with charge may lead towards gravitational collapse
of polytropic CO and can also be a source of cracking phenomenon.
The presence of cracking can be tested by means of methods provided by
Herrera \cite{21} and Gonzalez \cite{22}.
In this work, methods for the development of generalized LEe have been presented.
We observe that, in case \textbf{2} polytropes are not
physically viable due to violation of energy conditions.\\

\section*{Appendix}

\begin{eqnarray}\label{101}\notag
&&
3 \xi ^2 \theta ^n (\alpha _1+\alpha  \theta )+
\xi ^2 \theta  ^n (1+n \alpha _1+n \alpha  \theta  )+\\ \notag
&&
\frac{\alpha _4  \frac{d q}{d \xi}}{\xi  q  \alpha _3}
+\alpha  \xi ^3 \theta  ^n \frac{d \theta}{d \xi}+n \xi ^3 \theta  ^{-1+n}
(\alpha_1+\alpha  \theta  ) \frac{d \theta}{d \xi}
-\frac{\frac{d \alpha _4}{d \xi} }{\xi  \alpha _3}+\\ \notag
&&
\Big(-\frac{2 \theta  ^{-n} (\alpha ^2 \xi ^3 \alpha _3^2
\Delta  +2 \pi  q  P_{\text{rc}}^2 \frac{d q}{d \xi})}{\alpha  \xi ^2
P_{\text{rc}} \alpha _3^3 \alpha _2 }+\frac{\xi ^2
(\alpha _3+\frac{n \alpha _1}{\theta  })
\frac{d \theta}{d \xi}}{\alpha _3 \alpha _2 }\Big) \\ \notag
 &&
\Big(\frac{2 \alpha _3 v(\xi) }{\xi ^2}-\frac{2 \alpha _4 }
{\xi ^3}-\frac{2 \alpha _3 (\xi ^2 \theta  ^n (1+n \alpha _1+n \alpha
 \theta  )-\frac{\alpha _4 }{\xi ^2 \alpha _3}+\frac{\alpha _4
\frac{d q}{d \xi}}{\xi  q  \alpha _3})}{\xi }+\frac{\frac{d \alpha_4}{d \xi}}
{\xi ^2}\Big)+\\ \notag
 &&
\Big(1-\frac{2 \alpha _3 v(\xi) }{\xi }+\frac{\alpha _4 }{\xi ^2}\Big)
\Big(\frac{4 \theta  ^{-n} (\alpha ^2 \xi ^3 \alpha _3^2
\Delta  +2 \pi  q  P_{\text{rc}}^2 \frac{d q}{d \xi})}
{\alpha  \xi ^3 P_{\text{rc}} \alpha _3^3 \alpha _2 }+
\frac{2 \xi  (\alpha _3+\frac{n \alpha _1}{\theta  })
\frac{d \theta}{d \xi}}{\alpha _3 \alpha _2 }+\\ \notag
&&
\frac{2 n \theta  ^{-1-n} (\alpha ^2 \xi ^3 \alpha _3^2
\Delta  +2 \pi  q  P_{\text{rc}}^2 \frac{d q}{d \xi})
\frac{d \theta}{d \xi}}{\alpha
 \xi ^2 P_{\text{rc}} \alpha _3^3 \alpha _2 }-\frac{n
 \xi ^2 \alpha _1 (\frac{d \theta}{d \xi})^2}{\alpha _3
 \theta  ^2 \alpha _2 }+\\ \notag
 &&
\frac{2 \theta  ^{-n} (\alpha ^2 \xi ^3 \alpha _3^2
\Delta  +2 \pi  q  P_{\text{rc}}^2 \frac{d q}{d \xi})
\frac{d \alpha_2}{d\xi} }{\alpha
 \xi ^2 P_{\text{rc}} \alpha _3^3 \alpha _2^2}-\frac{\xi ^2
 (\alpha _3+\frac{n \alpha _1}{\theta  }) \frac{d \theta}{d \xi}
 \frac{d \alpha _2}{d \xi} }{\alpha _3 \alpha _2^2}-\\ \notag
  &&
\frac{2 \theta  ^{-n} (3 \alpha ^2 \xi ^2 \alpha _3^2 \Delta
+2 \pi  P_{\text{rc}}^2 (\frac{d q}{d \xi})^2+\alpha ^2 \xi ^3 \alpha _3^2
\frac{d\Delta}{d\xi} +2 \pi  q  P_{\text{rc}}^2
\frac{d^2 q}{d \xi^2})}{\alpha  \xi ^2 P_{\text{rc}} \alpha _3^3 \alpha _2 }+\\
&&
\frac{\xi ^2 (\alpha _3+\frac{n \alpha _1}{\theta})
\frac{d^2\theta}{d\xi^2} }{\alpha _3 \alpha _2 }\Big)=0,
\end{eqnarray}
\begin{eqnarray}\nonumber
&& \xi^2 \theta^n
+3 \xi^2 \theta^n (\alpha_1+\alpha \theta)
+\frac{\alpha_4 \frac{d q}{d \xi}}{\xi  q \alpha _3}
+\alpha  \xi^3 \theta^n \frac{d \theta}{d \xi}
+n \xi^3 \theta^{-1+n} (\alpha_1+\alpha  \theta )
\frac{d \theta}{d  \xi}\\ \notag
&& -\frac{\frac{d\alpha_4}{d\xi}}{\xi \alpha _3}+
\Big(- \frac{2 \theta^{-n} (\alpha^2 \xi^3 \alpha_3^2
\Delta+2 \pi  q P_{\text{rc}}^2 \frac{d q}{d \xi})}
{\alpha  \xi^2 P_{\text{rc}} \alpha_3^2 \alpha_5}
+\frac{\xi^2 (\alpha_3+\frac{n \alpha _1}{\theta})
\frac{d\theta}{d\xi}}{\alpha _3 \alpha_5}\Big)\\ \notag
&&\Big(\frac{2 \alpha _3 v }{\xi^2}
-\frac{2 \alpha _4}{\xi^3}
-\frac{2 \alpha _3 (\xi ^2 \theta^n-\frac{\alpha _4}
{\xi^2 \alpha _3}+\frac{\alpha _4 \frac{d q}{d \xi}}{\xi
q \alpha _3})}{\xi }
+\frac{\frac{d \alpha _4}{d\xi}}{\xi^2}\Big)
+\Big(1-\frac{2 \alpha _3 v }{\xi }+
\frac{\alpha _4}{\xi ^2}\Big)\\ \notag
&&\Big(\frac{4 \theta^{-n} (\alpha^2 \xi^3
\alpha _3^2 \Delta+2 \pi q P_{\text{rc}}^2 \frac{d q}{d\xi})}
{\alpha  \xi^3 P_{\text{rc}} \alpha _3^2 \alpha _5}
+\frac{2 \xi  (\alpha_3+\frac{n \alpha_1}{\theta})
\frac{d \theta}{d \xi}}{\alpha _3 \alpha _5}\\ \notag
&&+\frac{2 n \theta^{-1-n} (\alpha ^2 \xi ^3 \alpha _3^2
\Delta+2 \pi  q P_{\text{rc}}^2 \frac{d q}{d \xi}) \frac{d \theta}{d\xi}}
{\alpha \xi ^2 P_{\text{rc}} \alpha _3^2 \alpha _5}
-\frac{n \xi ^2 \alpha _1 (\frac{d\theta}{d\xi})^2}
{\alpha _3 \theta ^2 \alpha _5}\\ \notag
\end{eqnarray}
\begin{eqnarray}\nonumber
&&+\frac{2 \theta ^{-n} (\alpha ^2 \xi ^3 \alpha _3^2
\Delta +2 \pi  q P_{\text{rc}}^2 \frac{d q}{d\xi}) \frac{d\alpha _5}{d\xi}}
{\alpha \xi ^2 P_{\text{rc}} \alpha _3^2 \alpha _5^2}
-\frac{\xi ^2 (\alpha _3+\frac{n \alpha _1}{\theta })
\frac{d\theta}{d\xi} \frac{d\alpha_5}{d\xi}}{\alpha _3 \alpha _5^2}\\ \notag
&&-\frac{2 \theta ^{-n} (3 \alpha ^2 \xi ^2 \alpha _3^2 \Delta +2
\pi  P_{\text{rc}}^2 (\frac{d q}{d\xi})^2+\alpha ^2 \xi ^3 \alpha _3^2
\frac{d\Delta}{d\xi}+2 \pi  q P_{\text{rc}}^2 \frac{d^2 q}{d \xi^2})}
{\alpha  \xi ^2 P_{\text{rc}} \alpha _3^2 \alpha _5}\\
&&+\frac{\xi ^2 (\alpha _3+\frac{n \alpha _1}{\theta})
\frac{d^2\theta}{d\xi^2}}{\alpha _3 \alpha _5}\Big)=0,\label{102}
\end{eqnarray}
\begin{eqnarray}\label{103}\notag
&&3 \xi^2 \theta^n (\alpha _1+\alpha  \theta)+\xi ^2
\theta^n (1+n \alpha _1+n \alpha \theta )+\\ \notag
&&
\frac{\alpha _4  \frac{d q}{d \xi} }{\xi  q  \alpha _3}
+\alpha  \xi ^3 \theta^n \frac{d \theta}{d \xi} +n \xi^3 \theta^{-1+n}
(\alpha_1+\alpha \theta  ) \frac{d \theta}{d \xi} -\\ \notag
 &&
\frac{\frac{d \alpha _4}{d \xi} }{\xi  \alpha _3}+\Big
(-\frac{2 \xi  (1+n \alpha _1+n \alpha _1 \theta  )}
{\alpha _3 \alpha _2 }+\theta  ^{-n} \Big(-\frac{6
\alpha  v(\xi) }{\xi ^2 P_{\text{rc}} \alpha _3 \alpha _2 }+\\ \notag
&&
\frac{8 \alpha  \alpha _4 }{\xi ^3 P_{\text{rc}}
\alpha _3^2 \alpha _2 }+\frac{4 \pi  q  P_{\text{rc}}^2 \frac{d q}{d \xi} }{\alpha
 \xi ^2 \alpha _3^3 \alpha _2 }-\frac{4 \alpha
 \alpha _4  \frac{d q}{d \xi} }{\alpha _3^2 \alpha _2 })+\frac{\xi ^2
 (\alpha _3+\frac{n \alpha _1}{\theta  }) \frac{d \theta}
 {d \xi} }{\alpha _3 \alpha _2 }\Big) \\ \notag
&&
\Big(\frac{2 \alpha _3 v(\xi) }{\xi ^2}-\frac{2 \alpha _4 }
{\xi ^3}-\frac{2 \alpha _3 (\xi ^2 \theta  ^n (1+n \alpha _1+n
\alpha \theta  )-\frac{\alpha _4 }{\xi ^2 \alpha _3}
+\frac{\alpha _4  \frac{d q}{d \xi} }{\xi  q  \alpha _3})}{\xi }+
\frac{\frac{d \alpha _4}{d \xi} }{\xi ^2}\Big)+\\ \notag
&&
\Big(1-\frac{2 \alpha _3 v(\xi) }{\xi }+\frac{\alpha _4 }
{\xi ^2}\Big) \Big(-\frac{2 (1+n \alpha _1+n \alpha _1 \theta  )}
{\alpha _3 \alpha _2 }-\frac{2 n \xi  \alpha _1 \frac{d \theta}
{d \xi} }{\alpha _3 \alpha _2 }+\frac{2 \xi  (\alpha _3+\frac{n \alpha _1}
{\theta  }) \frac{d \theta}{d \xi} }{\alpha _3 \alpha _2 }-\\ \notag
&&
n \theta  ^{-1-n} \Big(-\frac{6 \alpha  v(\xi) }
{\xi ^2 P_{\text{rc}} \alpha _3 \alpha _2 }+\frac{8 \alpha  \alpha _4 }{\xi ^3
P_{\text{rc}}
\alpha _3^2 \alpha _2 }+\frac{4 \pi  q  P_{\text{rc}}^2
\frac{d q}{d \xi} }{\alpha  \xi ^2 \alpha _3^3 \alpha _2 }-\frac{4 \alpha
\alpha _4  \frac{d q}{d \xi} }{\alpha _3^2 \alpha _2 }\Big) \\ \notag
&&
\frac{d \theta}{d \xi} -\frac{n \xi ^2 \alpha _1
(\frac{d \theta}{d \xi})^2}{\alpha _3 \theta  ^2 \alpha _2 }
+\frac{2 \xi  (1+n \alpha _1+n \alpha _1
\theta   ) \frac{d \alpha _2}{d \xi} }{\alpha _3 \alpha _2 ^2}
-\frac{\xi ^2 (\alpha _3+\frac{n \alpha _1}{\theta  })
\frac{d \theta}{d \xi}  \frac{d \alpha _2}{d \xi} }
{\alpha _3 \alpha _2 {}^2}+\\ \notag
&&
\theta  ^{-n} \Big(\frac{12 \alpha  v(\xi) }{\xi ^3 P_{\text{rc}}
\alpha _3 \alpha _2 }-\frac{24 \alpha  \alpha _4 }{\xi ^4 P_{\text{rc}}
\alpha _3^2 \alpha _2 }-\frac{8 \pi  q  P_{\text{rc}}^2 \frac{d q}{d \xi} }
{\alpha  \xi ^3 \alpha _3^3 \alpha _2 }+\frac{4 \pi  P_{\text{rc}}^2
(\frac{d q}{d \xi})^2}{\alpha  \xi ^2 \alpha _3^3 \alpha _2 }-\\ \notag
&&
\frac{6 \alpha  (\xi ^2 \theta  ^n (1+n \alpha _1+n
\alpha  \theta  )-\frac{\alpha _4 }{\xi ^2 \alpha _3}+
\frac{\alpha _4  \frac{d q}{d \xi} }{\xi  q  \alpha _3})}
{\xi ^2 P_{\text{rc}} \alpha _3 \alpha _2 }+\frac{6 \alpha  v(\xi)
\frac{d \alpha _2}{d \xi}  }
{\xi^2 P_{\text{rc}} \alpha _3 \alpha _2 {}^2}-\\ \notag
&&
\frac{8 \alpha  \alpha _4  \frac{d \alpha _2}{d \xi} }{\xi ^3 P_{\text{rc}}
\alpha _3^2 \alpha _2 {}^2}-\frac{4 \pi  q  P_{\text{rc}}^2
\frac{d q}{d \xi}  \frac{d \alpha _2}{d \xi} }{\alpha  \xi ^2 \alpha _3^3
\alpha _2 {}^2}+\frac{4 \alpha  \alpha _4  \frac{d q}{d \xi}
\frac{d \alpha _2}{d \xi} }{\alpha _3^2
\alpha _2 ^2}+\frac{8 \alpha  \frac{d \alpha _4}{d \xi} }
{\xi ^3 P_{\text{rc}} \alpha _3^2 \alpha _2 }-\\
&&
\frac{4 \alpha  \frac{d q}{d \xi}  \frac{d \alpha _4}{d \xi} }
{\alpha _3^2 \alpha _2 }+\frac{4 \pi  q  P_{\text{rc}}^2
\frac{d^2 q}{d \xi^2} }{\alpha  \xi ^2
\alpha _3^3 \alpha _2 }-\frac{4 \alpha  \alpha _4  \frac{d^2 q}
{d \xi^2} }{\alpha _3^2 \alpha _2 }\Big)+\frac{\xi ^2 (\alpha _3+\frac{n
\alpha _1}{\theta  }) \frac{d^2 \theta}{d \xi^2} }{\alpha _3 \alpha _2 }\Big)=0,
\end{eqnarray}
\begin{eqnarray}\label{104}\notag
&&\xi ^2 \theta  ^n+3 \xi ^2 \theta  ^n
(\alpha _1+\alpha  \theta  )+\frac{\alpha _4  \frac{d q}{d \xi} }{\xi
 q  \alpha _3}+\\ \notag
&&
\alpha  \xi ^3 \theta  ^n \frac{d \theta}{d \xi} +n \xi ^3 \theta  ^{-1+n}
 (\alpha _1+\alpha  \theta  ) \frac{d \theta}{d \xi}
-\frac{\frac{d \alpha _4}{d \xi} }{\xi  \alpha _3}+\\ \notag
&&
\Big(-\frac{2 \xi }{\alpha _3 \alpha _5 }+\theta  ^{-n}
\Big(-\frac{6 v(\xi) }{\xi ^2 P_{\text{rc}} \alpha _3 \alpha _5 }+\frac{8
\alpha _4 }{\xi ^3 P_{\text{rc}} \alpha _3^2 \alpha _5 }+
\frac{4 \pi  q  P_{\text{rc}} \frac{d q}{d \xi} }{\alpha  \xi ^2 \alpha _3^3}-
\frac{4 \alpha _4  \frac{d q}{d \xi} }{\alpha _3^2 \alpha _5 }\Big)+\\ \notag
 &&
\frac{\xi ^2 (\alpha _3+\frac{n \alpha _1}{\theta  }) \frac{d \theta}{d \xi} }
{\alpha _3 \alpha _5 }\Big)
\Big(\frac{2 \alpha _3 v(\xi) }{\xi ^2}-\frac{2 \alpha _4 }
{\xi ^3}-\frac{2 \alpha _3 (\xi ^2 \theta  ^n-\frac{\alpha _4 }
{\xi^2 \alpha _3}+\frac{\alpha _4  \frac{d q}{d \xi} }{\xi  q  \alpha _3})}
{\xi }+\frac{\frac{d \alpha _4}{d \xi} }{\xi ^2}\Big)+\\ \notag
 &&
\Big(1-\frac{2 \alpha _3 v(\xi) }{\xi }+\frac{\alpha _4 }{\xi ^2}\Big)
\Big(-\frac{2}{\alpha _3 \alpha _5 }+\frac{2 \xi  (
\alpha _3+\frac{n \alpha _1}{\theta  }) \frac{d \theta}{d \xi} }
{\alpha _3 \alpha _5 }-\\ \notag
&&
n \theta  ^{-1-n} \Big(-\frac{6 v(\xi) }{\xi ^2 P_{\text{rc}}
\alpha _3 \alpha _5 }+\frac{8 \alpha _4 }{\xi ^3 P_{\text{rc}}
\alpha _3^2 \alpha _5 }+\frac{4 \pi  q  P_{\text{rc}}
\frac{d q}{d \xi} }{\alpha  \xi ^2 \alpha _3^3}-\frac{4 \alpha _4
\frac{d q}{d \xi} }{\alpha _3^2
\alpha _5 }\Big) \frac{d \theta}{d \xi} -\\ \notag
&&
\frac{n \xi ^2 \alpha _1 (\frac{d \theta}{d \xi}) ^2}{\alpha _3
\theta  ^2 \alpha _5 }+\frac{2 \xi  \frac{d \alpha _5}{d \xi} }
{\alpha _3 \alpha _5 ^2}-\frac{ \xi ^2 (\alpha _3+\frac{n \alpha _1}
{\theta  }) \frac{d \theta}{d \xi}  \frac{d \alpha _5}{d \xi} }
{\alpha _3 \alpha _5 {}^2}+\\ \notag &&
\theta  ^{-n} \Big(\frac{12 v(\xi) }{\xi ^3 P_{\text{rc}}
\alpha _3 \alpha _5 }-\frac{24 \alpha _4 }{\xi ^4 P_{\text{rc}}
\alpha _3^2 \alpha _5 }-\frac{8 \pi  q  P_{\text{rc}} \frac{d q}{d \xi} }
{\alpha  \xi ^3 \alpha _3^3}+\\ \notag &&
\frac{4 \pi  P_{\text{rc}} (\frac{d q}{d \xi}) ^2}{\alpha  \xi ^2
\alpha _3^3}-\frac{6 (\xi ^2 \theta  ^n-\frac{\alpha _4 }{\xi ^2 \alpha _3}
+\frac{\alpha _4  \frac{d q}{d \xi} }{\xi  q  \alpha _3})}{\xi ^2
P_{\text{rc}} \alpha _3 \alpha _5 }+\frac{8 \frac{d \alpha _4}{d \xi} }
{\xi ^3 P_{\text{rc}}
\alpha _3^2 \alpha _5 }-\\ \notag &&
\frac{4 \frac{d q}{d \xi}  \frac{d \alpha _4}{d \xi} }
{\alpha _3^2 \alpha _5 }+\frac{6 v(\xi)  \frac{d \alpha _5}{d \xi} }
{\xi ^2 P_{\text{rc}} \alpha _3 \alpha _5 {}^2}-\frac{8
\alpha _4  \frac{d \alpha _5}{d \xi} }{\xi ^3 P_{\text{rc}} \alpha _3^2
\alpha _5 {}^2}+\frac{4 \alpha _4  \frac{d q}{d \xi}  \frac{d \alpha _5}{d \xi} }
{\alpha _3^2 \alpha _5 {}^2}+\\
&&
\frac{4 \pi  q  P_{\text{rc}} \frac{d^2 q}{d \xi^2} }{\alpha  \xi ^2 \alpha _3^3}
-\frac{4 \alpha _4 \frac{d^2 q}{d \xi^2} }{\alpha _3^2 \alpha _5 }\Big)+
\frac{\xi ^2 (\alpha _3+\frac{n \alpha _1}{\theta  }) \frac{d^2 \theta}
{d \xi^2} }{\alpha _3 \alpha _5 }\Big)=0.
\end{eqnarray}

\end{document}